\documentclass[prb,twocolumn,superscriptaddress,amsmath,amssymb]{revtex4}
\usepackage{natbib}
\usepackage{graphicx}         
\usepackage{dcolumn}         
\usepackage{bm}                 

\begin{document}

\title{Phase-coherent transport in InN nanowires of various sizes}

\author{Ch. Bl\"omers}
\affiliation{Institute for Bio- and Nanosystems (IBN-1) and JARA
J\"ulich-Aachen Research Alliance, Research Centre J\"ulich GmbH,
52425 J\"ulich, Germany}

\author{Th. Sch\"apers}
\email{th.schaepers@fz-juelich.de} \affiliation{Institute for Bio-
and Nanosystems (IBN-1), JARA J\"ulich-Aachen Research Alliance,
and Virtual Institute of Spinelectronics (VISel), Research Centre
J\"ulich GmbH, 52425 J\"ulich, Germany}

\author{T. Richter}
\affiliation{Institute for Bio- and Nanosystems (IBN-1) and JARA
J\"ulich Aachen Research Alliance, Research Centre J\"ulich GmbH,
52425 J\"ulich, Germany}

\author{R. Calarco}
\affiliation{Institute for Bio- and Nanosystems (IBN-1) and JARA
J\"ulich-Aachen Research Alliance, Research Centre J\"ulich GmbH,
52425 J\"ulich, Germany}

\author{H. L\"uth}
\affiliation{Institute for Bio- and Nanosystems (IBN-1) and JARA
J\"ulich-Aachen Research Alliance, Research Centre J\"ulich GmbH,
52425 J\"ulich, Germany}

\author{M. Marso}
\affiliation{Institute for Bio- and Nanosystems (IBN-1) and JARA
J\"ulich-Aachen Research Alliance, Research Centre J\"ulich GmbH,
52425 J\"ulich, Germany}

\date{\today}

\hyphenation{InN}

\begin{abstract}
We investigate phase-coherent transport in InN nanowires of
various diameters and lengths. The nanowires were grown by means
of plasma-assisted molecular beam epitaxy. Information on the
phase-coherent transport is gained by analyzing the characteristic
fluctuation pattern in the magneto-conductance. For a magnetic
field oriented parallel to the wire axis we found that the
correlation field mainly depends on the wire cross section, while
the fluctuation amplitude is governed by the wire length. In
contrast, if the magnetic field is oriented perpendicularly, for
wires longer than approximately 200~nm the correlation field is
limited by the phase coherence length. Further insight into the
orientation dependence of the correlation field is gained by
measuring the conductance fluctuations at various tilt angles of
the magnetic field.
\end{abstract}

\maketitle


Semiconductor nanowires fabricated by a bottom-up
approach\cite{Thelander06,Lu06,Ikejiri07} have emerged as very
interesting systems not only for the design of future nanoscale
device structures\cite{Bjoerk02b,Bryllert06,Li06} but also to
address fundamental questions connected to strongly confined
systems. Regarding the latter, quantum dot
structures,\cite{DeFranceschi03,Fasth05a,Pfund06} single electron
pumps,\cite{Fuhrer07} or superconducting interference
devices\cite{VanDam06} have been realized. Many of the structures
cited above were fabricated by employing III-V semiconductors,
e.g. InAs or InP.\cite{Thelander06} Apart from these more
established materials, InN is particularly interesting for
nanowire growth because of its low energy band gap and its high
surface conductivity.\cite{Liang02,Chang05,Calarco07}

At low temperatures the transport properties of nanostructures are
affected by electron interference effects, i.e. weak localization,
the Aharonov--Bohm effect, or universal conductance
fluctuations.\cite{Beenakker91c,Lin02} The relevant length
parameter in this transport regime is the phase coherence length
$l_\phi$, that is the length over which phase-coherent transport
is maintained. In order to obtain information on $l_\phi$, the
analysis of conductance fluctuations is a very powerful
method.\cite{Umbach84,Stone85,Lee85,Altshuler85b,Lee87,Thornton87,Beenakker88a}
In fact, in InAs nanowires pronounced fluctuations in the
conductance have been observed and analyzed,
recently.\cite{Hansen05}

Here, we report on a detailed study of the conductance
fluctuations $\delta G$ measured in InN nanowires of various
sizes. Information on the phase-coherent transport is gained by
analyzing the average fluctuation amplitude and the correlation
field $B_c$. Special attention is drawn to the magnetic field
orientation with respect to the wire axis, since this allowed us
to change the relevant probe area for the detection of
phase-coherent transport.


The InN nanowires investigated here were grown without catalyst on
a Si (111) substrate by plasma-assisted
MBE.\cite{Calarco07,Stoica06a} The measured wires had a diameter
$d$ ranging from  42~nm to 130~nm. The typical wire length was
1~$\mu$m. From photoluminescence measurements an overall electron
concentration of about $5 \times 10^{18}$~cm$^{-3}$ was
determined.\cite{Stoica06a}

\begin{figure}
\includegraphics[width=1.0\columnwidth]{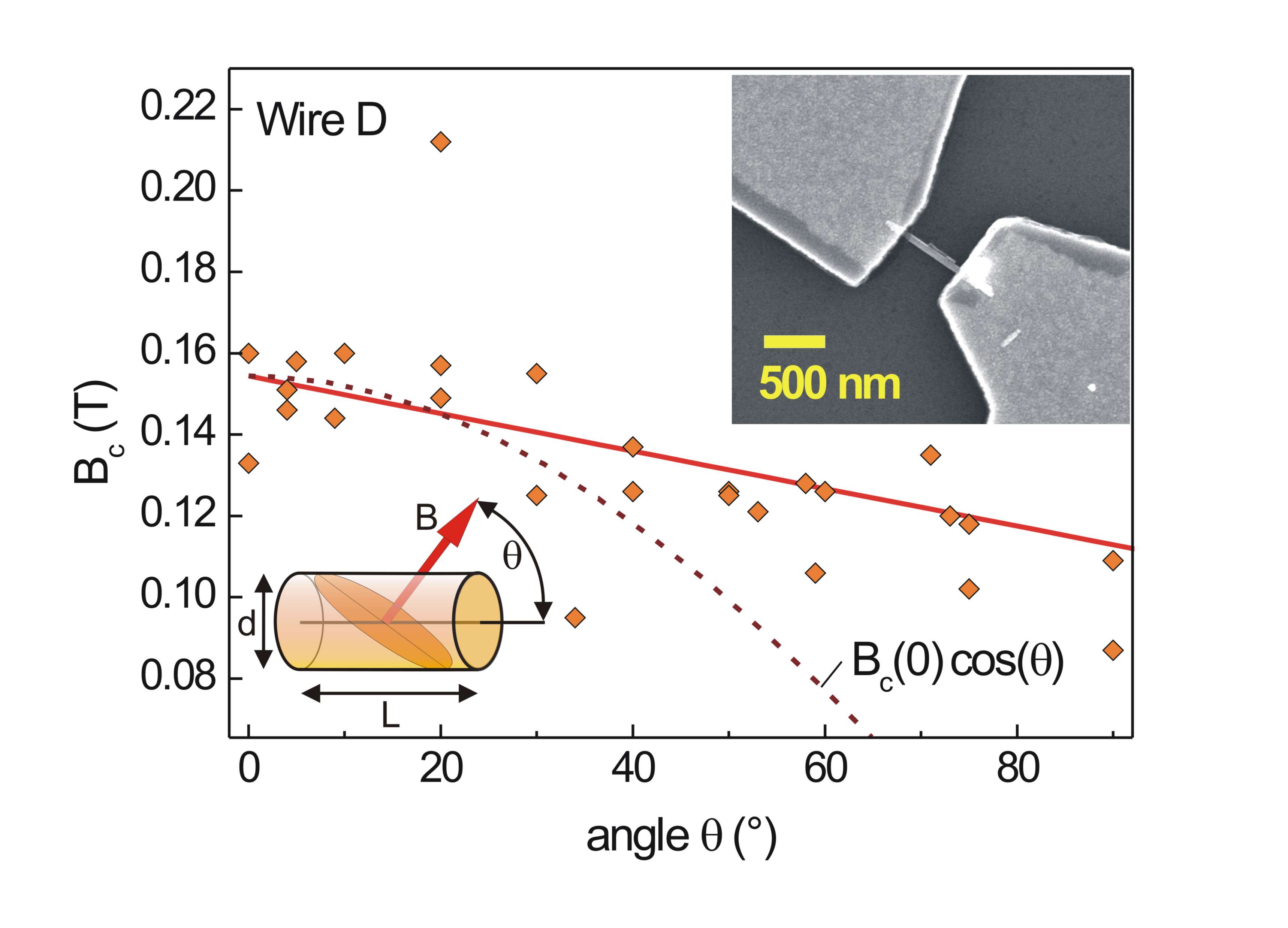}
\caption{Correlation field $B_c$ of wire~D as a function of the
angle $\theta$ between the wire axis and $B$. The solid line
represents a linear fit. The broken line corresponds to the
theoretically expected $B_c$ if phase-coherent transport is
assumed in the complete wire. The left-hand-side inset shows a
schematics of the geometrical situation. The right-hand-side inset
shows a micrograph of a 580-nm-long wire with a diameter of
66~nm.\label{Fig-Bc-vs-Angle}}
\end{figure}

For the samples used in the transport measurements, first, contact
pads and adjustment markers were defined on a SiO$_2$-covered Si
(100) wafer. Subsequently, the InN nanowires were placed on the
patterned substrate and contacted individually by Ti/Au
electrodes. Four wires labeled as A, B, C, and D will be discussed
in detail, below. Their parameters are summarized in
Table~\ref{Table1}. In order to improve the statistics, additional
wires which are not specifically labeled, were included in part of
the following analysis. A micrograph of a typical contacted wire
is depicted in Fig.~\ref{Fig-Bc-vs-Angle} (inset).
\begin{table}
\caption{Dimensions and characteristic parameters of the different
wires: Length $L$ (separation between the contacts), wire diameter
$d$, root-mean-square of the conductance fluctuations
$\mathrm{rms}(G)$, correlation field $B_c$. The latter two
parameters were determined for $B$ parallel to the wire axis.
 \label{Table1}}
 \begin{ruledtabular}
 \begin{tabular}{ccccccc}
Wire &  $L$ & $d$ & rms(G) & $B_c$\\
 &   (nm) &  (nm) & ($e^2/h$) & (T) \\ \colrule
A&  205 & 58 &1.35& 0.38\\
B&   580 & 66 &0.58& 0.22\\
C&  640 & 75 &0.52& 0.21\\
D&  530 & 130 &0.81& 0.15\\
 \end{tabular}
 \end{ruledtabular}
 \end{table}

The transport measurements were performed in a magnetic field
range from 0 to 10~T at a temperature of 0.6~K. In order to vary
the angle between the wire axis and the magnetic field $B$, the
samples were mounted in a rotating sample holder. The rotation
axis was oriented perpendicularly to the magnetic field and to the
wire axis. The magnetoresistance was measured by using a lock-in
technique with an ac bias current of 30~nA.

\begin{figure}
\includegraphics[width=1.0\columnwidth]{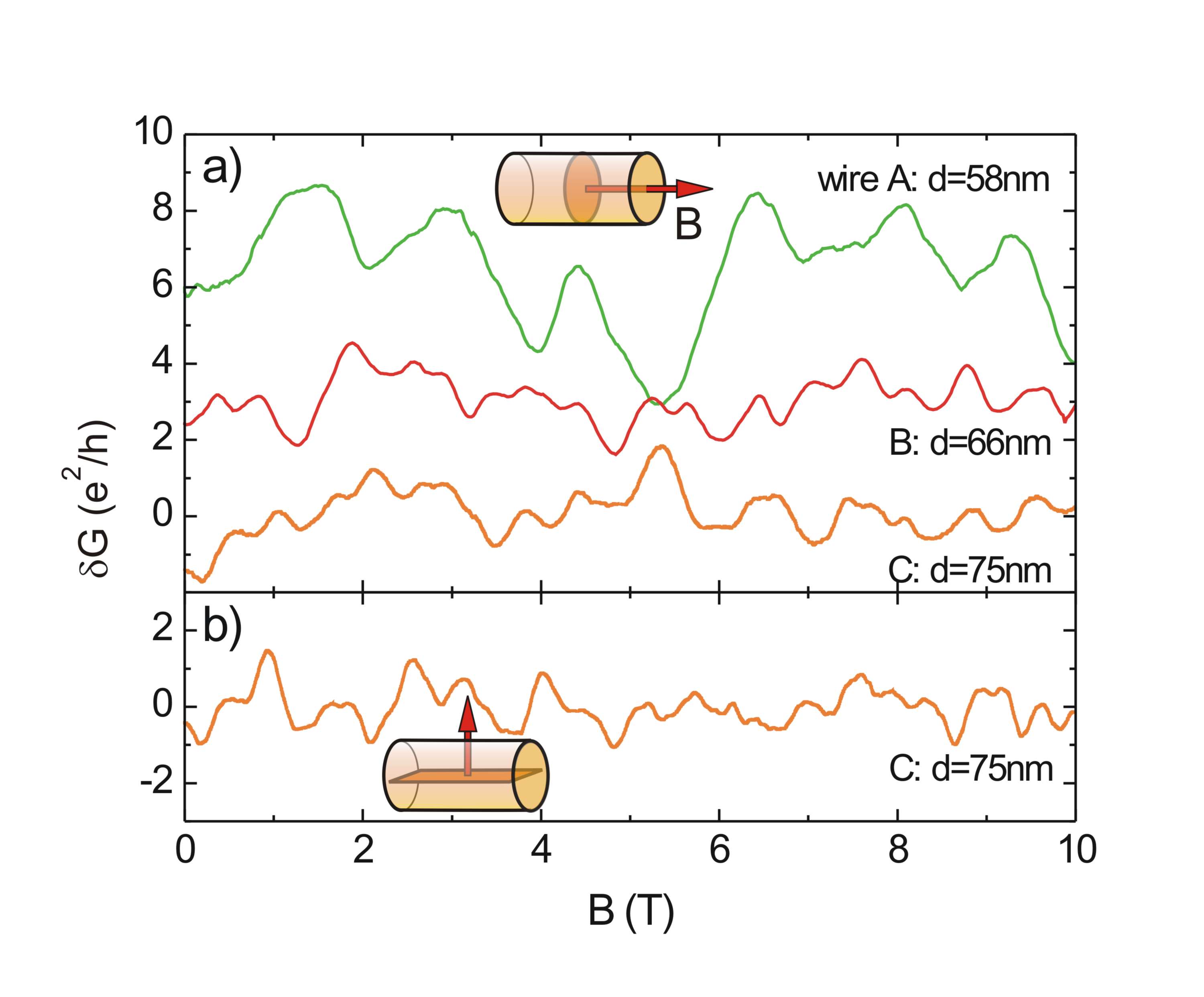}
\caption{(a) Conductance fluctuations normalized to $e^2/h$ for
wires with different length and diameter. The curves are offset
for clarity. As illustrated by the sketch, the magnetic field is
axially oriented. (b) Conductance fluctuations of wire~C with a
magnetic field oriented perpendicularly to the wire
axis.\label{Fig-compare-ucf}}
\end{figure}

The fluctuation pattern for nanowires with different dimensions
are depicted in Fig. \ref{Fig-compare-ucf}(a). Here, the
normalized conductance fluctuations $\delta G$ for wires A to C
comprising successively increasing diameters are plotted as a
function of the magnetic field $B$. The field was oriented
parallel to the wire axis. The measurements were performed up to a
relatively large field of 10~T. This is justified, since even at
10~T the estimated cyclotron diameter of 70~nm just begins to
become comparable to the wire diameter. The conductance variations
were determined by first subtracting the typical contact
resistance of $(330 \pm 50)\,\Omega$, and then converting the
resistance variations to conductance variations. It can clearly be
seen in Fig.~\ref{Fig-compare-ucf}(a), that for the narrowest and
shortest wire, i.e. wire~A, the conductance fluctuates with a
considerably larger amplitude than for the other two wires with
larger diameters and length. The parameter quantifying this
feature is the root-mean-square of the fluctuation amplitude
$\mathrm{rms}(G)$ defined by $\sqrt{\langle \delta G ^2\rangle}$.
Here, $\langle ... \rangle$ represents the average over the
magnetic field. For quasi one-dimensional systems where phase
coherence is maintained over the complete wire length it is
expected that $\mathrm{rms}(G)$ is in the order of
$e^2/h$.\cite{Lee85,Altshuler85b,Lee87} As one can infer from
Table~\ref{Table1}, for the shortest nanowire, i.e. wire A,
$\mathrm{rms}(G)$ falls within this limit. For the other two wires
the $\mathrm{rms}(G)$ values are smaller than $e^2/h$ (cf.
Table~\ref{Table1}). Thus, for these wires it can be concluded
that the phase coherence length $l_\phi$, is smaller than the wire
length $L$.

\begin{figure}
\includegraphics[width=1.0\columnwidth]{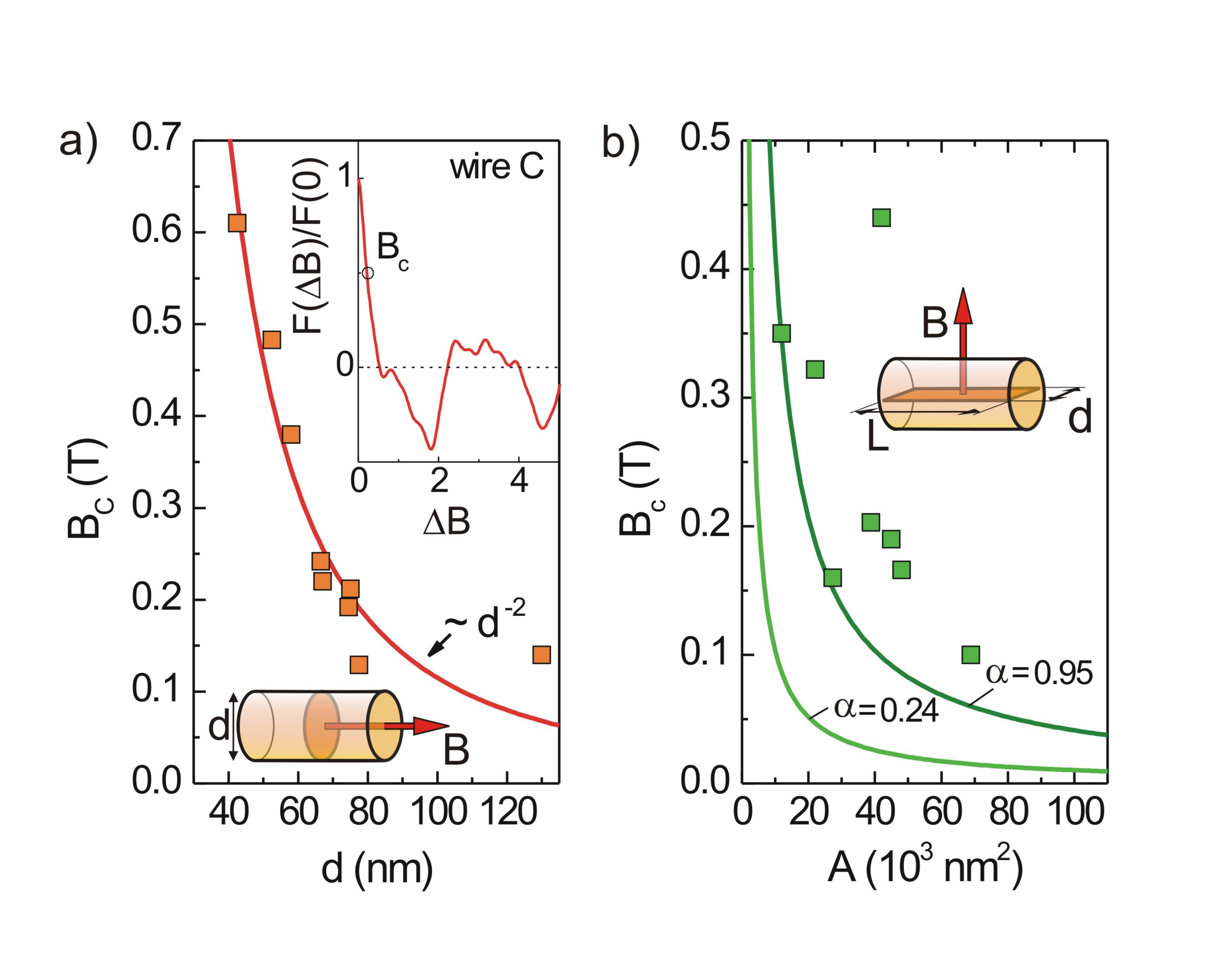}
\caption{(a) Correlation field $B_c$ as a function of the wire
diameter $d$. As illustrated in the schematics the magnetic field
$B$ was oriented axially. The solid lines corresponds to the
calculated correlation field. The inset shows $F(\Delta B)/F(0)$
for wire~C. (b) $B_c$ as a function of the maximum area $A=Ld$
(see schematics) of the wire. The magnetic field is oriented
perpendicular to the wire axis. The solid lines represents the
calculated lower boundary correlation fields assuming
$\alpha=0.95$ and 0.24, respectively. \label{Fig-BcvsDF1}}
\end{figure}

Beside $\mathrm{rms}(G)$, another important parameter is the
correlation field $B_c$, quantifying on which field scale the
conductance fluctuations take place. The correlation field is
extracted from the autocorrelation function of $\delta G$ defined
by $F(\Delta B)=\langle \delta G(B+\Delta B)\delta G (B)\rangle
$.\cite{Lee87} The magnetic field corresponding to half maximum of
the autocorrelation function $F(B_c)=\frac{1}{2} F(0)$ defines
$B_c$. The $B_c$ values of the measurements shown in
Fig.~\ref{Fig-compare-ucf} are listed in Table~\ref{Table1}.
Obviously, for wire~A, which has the smallest diameter, one finds
the largest value of $B_c$. In a semiclassical approach it is
expected that $B_c$ is inversely proportional to the maximum area
$A_\phi$ perpendicular to $B$ which is enclosed
phase-coherently:\cite{Lee85,Lee87,Beenakker88a}
\begin{equation}
B_c=\alpha \frac{\Phi_0}{A_\phi} \; . \label{Eq1}
\end{equation}
Here, $\alpha$ is a constant in the order of one and $\Phi_0=h/e$
the magnetic flux quantum. As long as phase coherence is
maintained along the complete circumference, $A_\phi$ is equal to
the wire cross section $\pi d^2/4$ and thus one expects $B_c
\propto 1/d^2$. The $B_c$ values given in Table~\ref{Table1}
follow this trend, i.e. becoming smaller for increasing diameter
$d$. As can be recognized in Fig.~\ref{Fig-BcvsDF1} (inset),
$F(\Delta B)$ also shows negative values at larger $\Delta B$.
This behavior can be attributed to the limited number of modes in
the wires, as it was observed previously for small size
semiconductor structures.\cite{Jalabert90,Bird96} However, as
discussed by Jalabert \emph{et al.}\cite{Jalabert90}, at small
fields $F(\Delta B)$ and thus $B_c$ being calculated fully quantum
mechanically correspond well to the semiclassical approximation.

In order to elucidate the dependence of $B_c$ on the wire diameter
in more detail, a larger number of wires was measured. As can be
seen in Fig.~\ref{Fig-BcvsDF1}(a), $B_c$ systematically decreases
with $d$. Leaving out wire~D which has the largest diameter, the
decrease of $B_c$ is well described by a $1/d^2$-dependence. As
mentioned above, for short wires ($L\approx 200$~nm) we found that
phase coherence is maintained over the complete length. This
length corresponds to a circumference of a wire with a diameter of
about 64 nm. Except of wire~D, $d$ is in the order of that value,
so that one can expect that phase coherence is maintained within
the complete cross section. For the parameter $\alpha$ we found a
value of 0.24, which is by a factor of 4 smaller than the
theoretically expected value of 0.95.\cite{Beenakker88a} Choosing
$\alpha=0.95$ would result in lower bound values of $B_c$ being
larger than all corresponding experimental values, which is
physically unreasonable. We attribute the discrepancy to the
different geometrical situation, i.e. for the latter a confined
two-dimensional electron gas with a perpendicularly oriented
magnetic field was considered,\cite{Beenakker88a} while in our
case the field is oriented parallel to the wire axis. In addition,
an inhomogeneous carrier distribution within the cross section,
e.g. due to a carrier accumulation at the surface,\cite{Mahboob04}
can also result in a disagreement between experiment and
theoretical model. As can be seen in Fig.~\ref{Fig-BcvsDF1}(a)
(inset), the data point of the wire with the largest diameter of
130~nm, i.e. wire~D, is found above the calculated curve. This
indicates that presumably for this sample, $A_\phi$ is slightly
smaller than the wire cross section.

Next, we will focus on measurements of $\delta G$ with a magnetic
field oriented perpendicular to the wire axis. As a typical
example, $\delta G$ of wire~C is shown in
Fig.~\ref{Fig-compare-ucf}(b). Here, a correlation field of
$0.17$~T was extracted, which is smaller than the value of
corresponding measurements with $B$ parallel to the wire axis
[c.f. Fig.~\ref{Fig-compare-ucf}(a) and Table~\ref{Table1}]. The
smaller value of $B_c$ can be attributed to the effect that now
the relevant area for magnetic flux-induced interference effects
is no longer limited by the relatively small circular cross
section but rather by a larger area within the rectangle defined
by $L$ and $d$, as illustrated by the schematics in
Fig.~\ref{Fig-BcvsDF1}(b).

In Fig.~\ref{Fig-BcvsDF1}(b) the $B_c$ values of various wires are
plotted as a function of the maximum area $A_{max}=Ld$ penetrated
by the magnetic field. As a reference, the calculated curve using
Eq.~(\ref{Eq1}) and assuming $A_\phi=A_{max}$ are also plotted. It
can be seen that the $B_c$ values of two wires with small areas,
including wire~A, match to the theoretically expected ones if one
takes $\alpha=0.95$, as given by Beenakker and van
Houten.\cite{Beenakker88a} This corresponds to the case of
phase-coherent transport across the complete wire, as it was, in
case of wire~A, already concluded from the $\mathrm{rms}(G)$
analysis. For all other wires the $B_c$ values are above the
theoretically expected curve, corresponding to the case
$A_\phi<A_{max}$. At this point, one might argue that for $B$
oriented along the wire axis a better agreement is found for
$\alpha=0.24$. However, as can be seen in
Fig.~\ref{Fig-BcvsDF1}(b), if one assumes $\alpha=0.24$ all
experimental values are above the calculated curve, i.e.
$A_\phi<A_{max}$. This does not agree with the observation that
for short wires $\mathrm{rms}(G)$ is in the order of $e^2/h$. We
attribute the difference between the appropriate $\alpha$ values
for different field orientations to the different character of the
relevant area penetrated by the magnetic flux, e.g. due to carrier
accumulation at the surface.

\begin{figure}
\includegraphics[width=1.0\columnwidth]{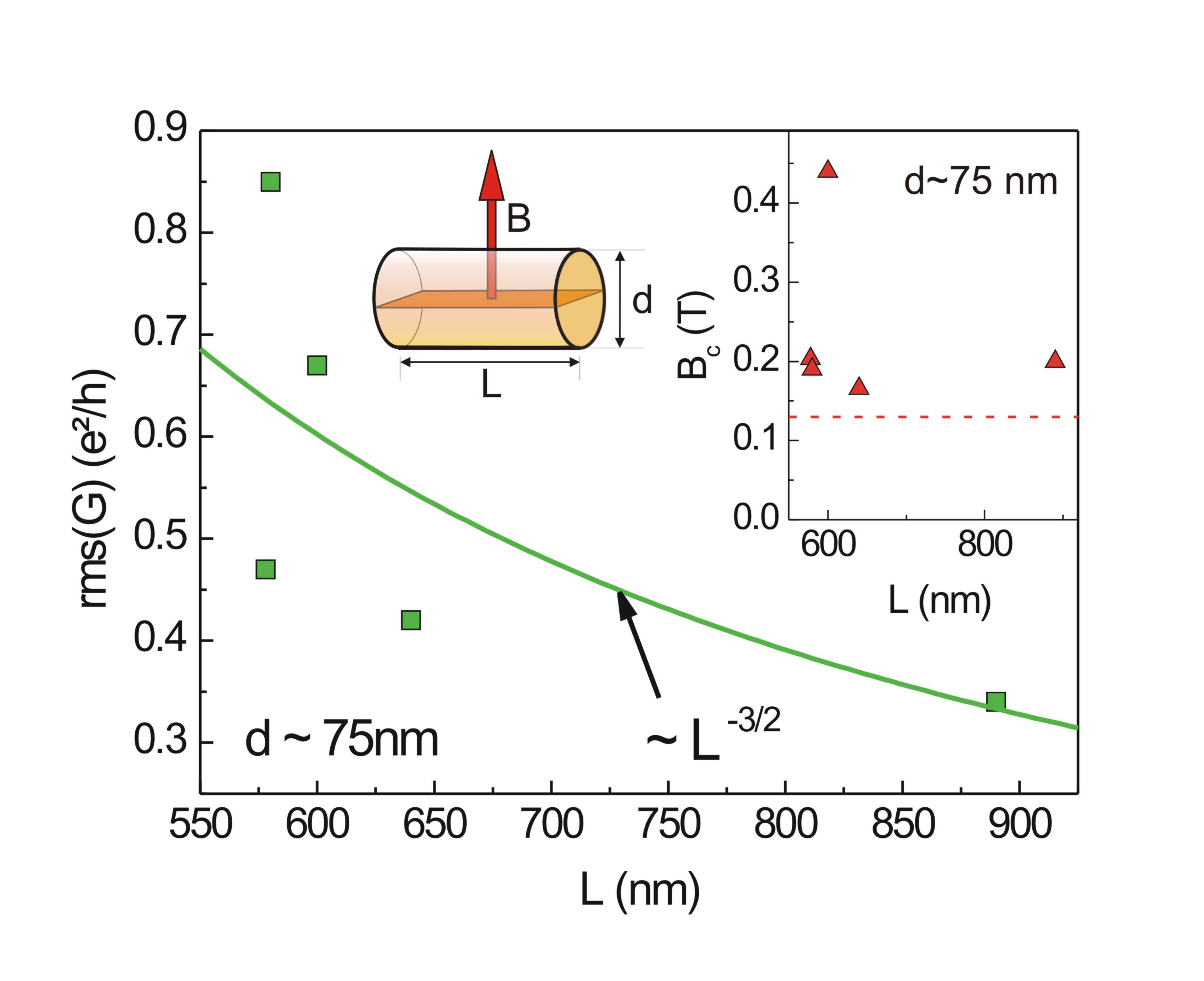}
\caption{$\mathrm{rms}(G)$ for wires with a diameter of $(75 \pm
5)$~nm as a function of wire length $L$ (square). The magnetic
field is oriented perpendicular to the wire axis. The calculated
decrease of $\mathrm{rms}(G)$ proportional to $L^{-3/2}$ is
plotted as solid line. The inset shows $B_c$ vs. $L$ for wires
with $d \approx 75$~nm. The dashed line corresponds to the
calculated value of $B_c$ assuming
$l_\phi=430$~nm.\label{Fig-rms-75nm}}
\end{figure}

Beside $B_c$ we also analyzed the fluctuation amplitude for five
different wires with $B$ oriented perpendicular to the wire axis.
Only wires with comparable diameters of ($75 \pm 5$)~nm were
chosen, here. It can be seen in Fig.~\ref{Fig-rms-75nm} that
$\mathrm{rms}(G)$ tends to decrease with increasing wire length
$L$.
From the previous discussion of $B_c$ it was concluded that for
long wires, as it is the case here, $l_\phi<L$. In this regime
$\mathrm{rms}(G)$ is expected to depend on $L$
as\cite{Lee87,Beenakker88a}
\begin{equation}
\mathrm{rms}(G)=\beta \frac{e^2}{h} \left(\frac{l_\phi}{L}
\right)^{3/2} \; , \label{Eq2}
\end{equation}
with $\beta$ in the order of one. The above expression is valid as
long as the thermal diffusion length $l_T=\sqrt{\hbar
\mathcal{D}/k_BT}$, is larger than $l_\phi$. Here, $\mathcal{D}$
is the diffusion constant. From our transport data we estimated
$l_T \approx 600$~nm at $T=0.6$~K. As can be seen in
Fig.~\ref{Fig-rms-75nm}, the available experimental data points
roughly follow the trend of the calculated curve using
Eq.~(\ref{Eq2}) and assuming $l_\phi=430$~nm and $\beta=1$. For
the limit $l_\phi < L$, a correlation field according to $B_c=0.95
\Phi_0/d \l_\phi$ is expected.\cite{Lee87} As confirmed in
Fig.~\ref{Fig-BcvsDF1}(b), most experimental values of $B_c$ are
close to the calculated one.

If one compares the $\mathrm{rms}(G)$ values for wires with $d
\approx 75$~nm and $B$ oriented axially (not shown here) with the
corresponding values for $B$ oriented perpendicularly, one finds,
that both are in the same range. Thus it can be concluded that the
fluctuation amplitude does not significantly depend on the
magnetic field orientation. This is in contrast to the correlation
field, where one finds a systematic dependence on the orientation
of $B$.

In order to discuss the latter aspect in more detail the
correlation field was studied for various tilt angles $\theta$ of
the magnetic field. Figure~\ref{Fig-Bc-vs-Angle} shows $B_c$ of
sample~D if $\theta$ is increased from $0^\circ$ to $90^\circ$.
The inset in Fig.~\ref{Fig-Bc-vs-Angle} illustrates how $\theta$
is defined.
Obviously, $B_c$ decreases with increasing tilt angle $\theta$. As
explained above, the value of $B_c$ is a measure of the maximum
area normal to $B$, which is enclosed phase-coherently by the
electron waves in the wire [see Fig.~\ref{Fig-Bc-vs-Angle}
(schematics)]. As long as $\theta \leq \arctan (L/d)$, this
maximum area is given by $A(\theta)=\pi d^2/4\cos{\theta}$. The
expected $\theta$-dependence of the correlation field is then
given by $B_c(\theta)$=$B_c(0) \cos(\theta)$, with $B_c(0)$ the
correlation field at $\theta=0$. As can be seen in
Fig.~\ref{Fig-Bc-vs-Angle}, the calculated correlation field
$B_c$, corresponding to fully phase-coherent transport, decreases
much faster with increasing $\theta$ than the experimentally
determined values. The experimental situation is better described
by a linear decrease. As it was discussed above, at $\theta=0$ one
can assume that the area enclosed phase-coherently is equal to
$A(0)$. However, if the tilt angle is increased the maximum wire
cross section $A(\theta)$ presumably becomes larger than $A_\phi$,
resulting in a much smaller decrease of $B_c$ than theoretically
expected for fully phase-coherent transport. In addition, as
pointed out above, the different tilt angles result in an
angle-dependent parameter $\alpha$. This is supported by the
measurements of $B_c$ for $B$ parallel and perpendicular to the
wire axis, where different values for $\alpha$ were determined,
respectively.


In conclusion, the conductance fluctuations of InN nanowires with
various lengths and diameters were investigated. We found that for
an axially oriented magnetic field the correlation field $B_c$ and
thus the area where phase-coherent transport is maintained is
limited by the wire cross section perpendicular to $B$. In
contrast, $\mathrm{rms}(G)$ decreases with the wire length, since
this quantity also depends on the propagation of the electron
waves along the wire axis. If the magnetic field is oriented
perpendicularly we found that for long wires $B_c$ is limited by
$l_\phi$ rather than by the length $L$. Our investigations
demonstrate that phase-coherent transport can be maintained in InN
nanowires, which is an important prerequisite for the design of
quantum device structures based on this material system.

\end{document}